# Synergies between Federated Foundation Models and Smart Power Grids

Seyyedali Hosseinalipour, *Senior Member, IEEE,* Shimiao Li, *Member, IEEE,*
Adedoyin Inaolaji, *Member, IEEE,* Filippo Malandra, *Member, IEEE,*
Luis Herrera, *Senior Member, IEEE,* and Nicholas Mastronarde, *Senior Member, IEEE*

*Abstract*—The recent emergence of large language models (LLMs) such as GPT-3 has marked a significant paradigm shift in machine learning. Trained on massive corpora of data, these models demonstrate remarkable capabilities in language understanding, generation, summarization, and reasoning, transforming how intelligent systems process and interact with human language. Although LLMs may still seem like a recent breakthrough, the field is already witnessing the rise of a new and more general category: multi-modal, multi-task foundation models (M3T FMs). These models go beyond language and can process heterogeneous data types/modalities, such as time-series measurements, audio, imagery, tabular records, and unstructured logs, while supporting a broad range of downstream tasks spanning forecasting, classification, control, and retrieval. When combined with federated learning (FL), they give rise to M3T Federated Foundation Models (FedFMs): a highly recent and largely unexplored class of models that enable scalable, privacy-preserving model training/fine-tuning across distributed data sources. In this paper, we take one of the first steps toward introducing these models to the power systems research community by offering a bidirectional perspective: *(i) M3T FedFMs for smart grids* and *(ii) smart grids for FedFMs*. In the former, we explore how M3T FedFMs can enhance key grid functions, such as load/demand forecasting and fault detection, by learning from distributed, heterogeneous data available at the grid edge in a privacy-preserving manner. In the latter, we investigate how the constraints and structure of smart grids, spanning energy, communication, and regulatory dimensions, shape the design, training, and deployment of M3T FedFMs.

## I. Introduction

Smart grids are modernized power systems that integrate sensing, communication, and control technologies to manage electricity generation, transmission, distribution, and consumption with greater efficiency and resilience than conventional power grids. As these systems deploy more sensors, advanced meters, and automation tools, they *produce large volumes of data*, ranging from time-series measurements and geo-spatial records to textual logs and imagery. On the one hand, machine learning (ML) models are being increasingly adopted to make sense of this data and drive key system operations such as load forecasting, fault detection, voltage regulation, and distributed energy resource (DER) optimization. On the other hand, smart grids also underpin the infrastructure that powers ML itself: they generate the data, provide the energy supply for edge and cloud computation, and define the physical and operational constraints under which ML models are deployed. This gives rise to a two-way paradigm: *ML for smart grids*, where learning models enable intelligent control and decision-making for smart grid operations, and *smart grids for ML*, where the smart grid serves as both a deployment environment and a catalyst for advancing robust, scalable ML systems.

This two-way paradigm places growing demands on the underlying ML models used in smart grid applications. Conventional ML models, such as convolutional neural networks (CNNs) or recurrent neural networks (RNNs), are typically designed for specific tasks and input modalities. These models are often trained independently from each other for narrow use cases (e.g., ML tasks for load forecasting). However, with smart grid data spanning imagery, time-series, tabular logs, and text, maintaining a fragmented set of task-specific models becomes inefficient and unsustainable. This issue calls for a shift toward the adoption of *foundation models (FMs)* in smart grid settings: general-purpose architectures often pre-trained on broad data domains and designed to support multiple ML tasks through modular heads or lightweight techniques.

FMs emerged as a dominant paradigm in ML following the rapid advancement of *large language models (LLMs)*, such as GPT-3, which demonstrated unprecedented performance across a wide range of natural language processing tasks. These models popularized the concept of pre-training on massive, diverse datasets followed by fine-tuning for downstream applications, a strategy that proved highly effective at scale. Motivated by their success, recent studies have begun exploring LLMs in power grid contexts, leveraging their ability to process and generate structured text for applications like outage report summarization, maintenance log analysis, and operator decision support.

Although LLMs may still feel cutting-edge, they are already considered to be the first wave in the recent evolution of ML models focused primarily on language and text-based ML tasks (e.g., question answering, summarization, dialogue generation, code completion, and document classification). In particular, a more recent surge has given rise to *multi-modal, multi-task foundation models (M3T FMs)*: architectures capable of jointly learning from multiple data modalities (e.g., time-series data, images, tabular records, and text), while supporting a range of downstream ML tasks. This evolution marks a significant leap, aligning with the structure and demands of modern smart grids where multiple modalities of data are collected across the *grid edge* for various downstream ML tasks. However, since they have only recently emerged in models such as GPT-4, Flamingo, and Gemini, M3T FMs remain largely unexplored in smart power grid applications, with only a few early efforts beginning to examine their role.

Despite their immerse potential, M3T FMs face a critical challenge when applied to real-world domains such as smart grids: lack of access to sufficient and diverse training data. Specifically, these models are inherently data-hungry, relying on large-scale, multi-modal datasets to capture generalizable patterns across tasks. However, in practice, much of the relevant data in power systems is *siloed* across distinct entities, ranging from local utilities and municipal microgrids to private grid operators and regional transmission organizations. Here, operational constraints, regulatory barriers, competitive interests, and privacy requirements often prevent these stakeholders from sharing their raw data. As a result, the centralized training/fine-tuning pipelines typically used for M3T FMs are often infeasible. Addressing this challenge requires rethinking how M3T FMs are trained, specifically, by adopting distributed learning frameworks that respect data locality. Among these frameworks, federated learning (FL) is the most prominent approach, enabling collaborative model training across decentralized datasets without revealing sensitive raw data. Notably, the convergence of M3T FMs with FL has led to the emergence of a new and rapidly evolving class of models known as *M3T Federated Foundation Models (FedFMs)*. Nevertheless, despite their potential, M3T FedFMs remain unexplored in the smart grid domain due to their extremely recent origin, forming the main motivation behind this work.

In this paper, we explore the dual paradigm of *M3T FedFMs for power grids* and *power grids for M3T FedFMs*. Specifically, the first direction (i.e., M3T FedFMs for power grids) examines how these models can serve as unified, scalable learning frameworks that ingest diverse grid data (e.g., sensor streams, imagery, operational logs) and support a wide range of downstream tasks such as forecasting, anomaly detection, fault localization, and control. Also, the second direction (i.e., power grids for M3T FedFMs) flips the perspective, asking how the structure, constraints, and data ecosystem of power grids can define and also affect the development of M3T FedFMs. Together, this dual lens reveals both the transformative potential of M3T FedFMs in enabling intelligent grid operations and the fundamental role that smart grids can play in shaping the future of M3T FedFMs.

## II. Multi-Modal Multi-Task Federated Foundation Models (M3T FedFMs)

In the following, we provide background on the core concepts of FL, multi-modal and multi-task learning, and M3T FMs.

### A. Federated Learning (FL)

FL is a distributed ML paradigm designed to enable collaborative model training across multiple data holders (called *nodes*) without requiring raw data transfer across the system. Conventional FL typically follows a four-step iterative procedure: *(i) local training,* where each participating node (e.g., utility or substation) trains a local model on its private dataset; *(ii) model upload,* where the locally updated model/gradient parameters are sent to a central server; *(iii) model aggregation,* where the server combines the updates (often via weighted averaging) to produce a global model; and *(iv) model broadcast,* where the global model is broadcast back to the participants for the next round of local training. This cycle repeats until the global model converges or reaches a desired loss/accuracy. By keeping raw data local at each participating node and only exchanging model/gradient parameters, FL enables privacy-preserving distributed ML, making it particularly attractive for domains such as power systems where data sensitivity and regulatory compliance are critical. As a result, FL has been applied across various smart grid scenarios, including load forecasting across utility regions, fault detection using distributed sensor data, collaborative energy management among microgrids, and demand response modeling. Nevertheless, most of the literature focuses on training conventional ML models (e.g., CNNs and RNNs) that lack the generalizability and adaptability of FMs.

### B. Multi-Modal and Multi-Task Learning Across Smart Grids

Smart grids generate rich *multi-modal data*, reflecting the complexity of their physical infrastructure and operational workflows, which can be broadly categorized into five key modalities: *(i) Time-series* data forms the backbone of grid monitoring, capturing sequential measurements such as voltage, current, frequency, and load profiles from phasor measurement units (PMUs), smart meters, and battery management systems. Complementing this are *(ii) structured tabular data*, such as supervisory control and data acquisition (SCADA) logs, maintenance schedules, DER metadata, and market transactions, which collectively record system states, events, and asset attributes. *(iii) Textual data*, which can be human-generated, includes outage reports, maintenance logs, and operator notes, offering insights into past events and decision-making processes. *(iv) Visual data* in the form of drone imagery, thermal scans, and surveillance video feeds has become crucial for identifying physical anomalies such as vegetation overgrowth, equipment degradation, or heat stress. Finally, *(v) environmental and contextual data modalities*, including weather conditions, geographic information, time-of-day indicators, and terrain data, provide essential external signals that influence demand patterns, equipment performance, and outage risks.

In parallel with its multi-modal data landscape, the smart grid is inherently a *multi-task environment*, where the corresponding ML tasks can be broadly categorized into four classes: *(i) forecasting tasks*, such as load prediction, renewable generation estimation, and electricity price forecasting; *(ii) classification tasks*, including fault type detection, equipment status identification, and cybersecurity threat recognition; *(iii) regression tasks*, such as state estimation, voltage margin estimation, line impedance prediction, and thermal loading analysis; *(iv) control, optimization, and decision-making tasks*, including inverter tuning, load shedding, demand response coordination, energy dispatch, topology reconfiguration, and scheduling of DERs.

Several studies highlight that treating the aforementioned modalities and tasks in isolation misses critical inter-modality and inter-task dependencies, advocating for *multi-modal learning (MML) and multi-task learning (MTL)*. In MML, models jointly leverage diverse data types to uncover correlations among the modalities (e.g, through using a unified neural network with dedicated encoders for different modalities, the outputs of which are fused together via various fusion techniques such as *attention* mechanisms) and improve task performance beyond what single-modality models can achieve. This approach has demonstrated superior results in fault diagnosis, load monitoring, and predictive maintenance, where the use of complementary modalities are shown to enhance model performance. Also, MTL enables training of related tasks using shared representations (e.g, through using a unified neural network with dedicated outputs/heads for different tasks and shared layers that capture inter-task correlations). MTL has outperformed single-task models in diverse applications such as non-intrusive load monitoring, fault-detection, and solar forecasting and shown to enable simultaneous fault type, location, and resistance estimation.

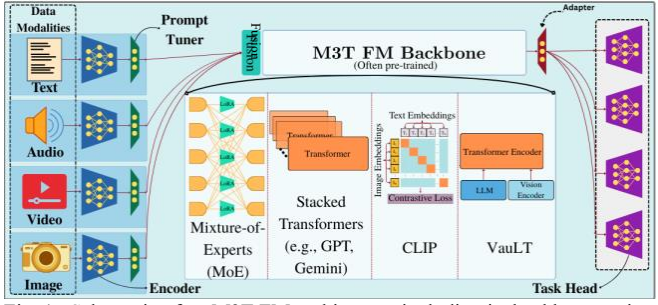

Fig. 1. Schematic of an M3T FM architecture including its backbone variants.

Despite these advancements, a common thread across much of the existing literature is that studies tend to focus on either multi-modal learning for a single task (e.g., combining imagery and sensor data for fault detection) or multi-task learning using a single data modality (e.g., performing forecasting and classification over time-series data). This separation fails to fully exploit the rich cross-modal and cross-task dependencies present in smart grid environments. M3T FMs offer a compelling pathway to bridge this gap. Moreover, M3T FMs elevate this paradigm by introducing *generative* capabilities, a shift beyond traditional discriminative ML tasks such as classification or regression. Specifically, these models can synthesize reports, summarize multi-source incident logs, generate future grid state trajectories, or even produce counterfactual explanations, thereby opening new possibilities for enhanced situational awareness and optimal decision-making across both operational and planning horizons. When combined with FL capabilities, these models become even more powerful: they can learn collaboratively across geo-dispersed utility nodes without compromising data privacy and adapt to regional variations in data modalities and ML tasks. Motivated by this, we next delve into the structure of M3T FMs and M3T FedFMs.

## III. M3T FedFMs for Smart Grids: Enabling New Frontiers of Distributed Intelligence

In this section, we focus on examining the integration of M3T FedFM training/fine-tuning within power grid infrastructures and highlight the unique challenges and considerations associated with their training and deployment in this domain.

### A. M3T FedFMs Integration in Smart Grids

In M3T FedFMs, each participating node (e.g., a substation, microgrid controller, or regional operator) maintains a local instance of an M3T FM and trains/fine-tunes it using its own private data. These underlying M3T FMs (depicted in Fig. 1) are architecturally composed of several coordinated modules: *(i) modality-specific encoders* which transform the input data modalities into processed embeddings/representations, *(ii) a shared backbone* that fuses information across embeddings, and *(iii) task-specific heads* designed for mapping the output of the shared backbone to ML tasks of interest (e.g., outage report summarization, load forecasting, and anomaly detection). To improve scalability and computational efficiency, some M3T FMs also incorporate Mixture-of-Experts (MoEs): layers composed of multiple parallel expert sub-networks (e.g., neural networks or transformers). During training or inference, only a small subset of these experts is activated per input instance. In addition to reducing the computational burden, this conditional routing mechanism allows the model to dynamically specialize pathways for different data types or task contexts, enabling improved performance in multi-task, multi-domain settings.

In the M3T FedFM setting (see Fig. 2), each node locally trains/fine-tunes its M3T FM, where the updates are periodically aggregated (e.g., using weighted averaging or more advanced aggregation schemes) to form a shared global model. Rather than fine-tuning the entire model end-to-end, nodes often adapt only the components relevant to their available data and operational tasks. For instance, a utility focused on forecasting may update the temporal encoder and forecasting task head, while a substation performing fault analysis may fine-tune the image encoder and classification head. To further reduce computational and communication overhead, many M3T FedFMs employ lightweight fine-tuning strategies, such as adapter tuning, prompt tuning, or LoRA (Low-Rank Adaptation), which update a small set of inserted trainable parameters while keeping most of the pre-trained model frozen.

### B. M3T FedFMs Applications in Smart Grids

The integration of M3T FedFMs into smart grids enables more than just predictive modeling, it introduces generative capabilities that can synthesize reports, simulate future scenarios, and produce actionable recommendations by reasoning over multi-modal data streams. Below, we present three representative applications that can leverage the unique strengths of M3T FedFMs.

*1) Proactive Fault Detection, Localization, and Incident Report Generation:* Fault management in modern smart grids is a prototypical multi-modal, multi-task setting. In this application, time-series measurements from PMUs and smart meters (e.g., voltage, current, frequency waveforms) offer precise electrical signatures of anomalies; thermal and optical imagery from drones reveal visual indicators of component degradation; and unstructured textual logs from maintenance crews capture contextual and causal details not present in sensor data. M3T FedFMs can unify these heterogeneous input data types through modality-specific encoders and a shared backbone that jointly learns electrical, visual, and textual representations. This enables simultaneous execution of multiple downstream tasks: (i) fault type classification, (ii) geospatial localization of the fault, and (iii) regression-based estimation of fault severity. Beyond these discriminative capabilities, the generative capabilities of M3T FedFMs can introduce a qualitative leap: given the fused multi-modal evidence, the model can synthesize operator-ready incident reports that articulate the probable cause (e.g., salt corrosion on coastal equipment, conductor sag due to thermal overload), summarize supporting sensor and image evidence, and recommend prioritized mitigation steps.

*2) Renewable Generation Forecasting and Generative Scenario Simulation:* Renewable energy integration exemplifies the need for both multi-modal and multi-task learning in smart grids. In this application, time-series data from photovoltaic inverters and wind turbines (e.g., power output, rotor speed) capture the instantaneous operational state; satellite and sky imagery provide visual indicators of cloud cover and atmospheric conditions; and structured weather forecasts and environmental measurements (temperature, wind speed, humidity) add predictive context. M3T FedFMs can use these diverse data modalities in a unified learning architecture, where temporal encoders extract patterns from inverter signals, vision encoders identify cloud motion and density from imagery, and tabular encoders process structured meteorological data. The shared backbone can then fuse these embeddings to support multiple downstream tasks: (i) short-term generation forecasting, (ii) ramp event detection, and (iii) quantification of forecast uncertainty. Also, the generative capabilities of M3T FedFMs extend this capability by simulating "what-if" scenarios: given a hypothetical weather shift (e.g., rapid cloud buildup from a coastal front or sudden temperature drop in a high-altitude wind farm), the model can potentially generate multiple plausible power generation trajectories, each annotated with probability estimates and potential grid impacts. These generative simulations allow operators to explore alternative dispatch strategies, enabling proactive decision-making.

*3) DER Coordination with Generative Control Policies:* Coordinating DERs such as battery storage systems, rooftop solar, and controllable loads is inherently a multi-modal, multi-task challenge. In this application, SCADA tabular records and DER operational logs (e.g., state-of-charge, inverter settings) provide structured understanding of the system state; real-time sensor streams supply granular electrical measurements (voltage, frequency, load); and market data offers dynamic economic signals (electricity prices, demand response events). M3T FedFMs can integrate these geo-dispersed data modalities through dedicated encoders whose outputs are fused in a shared backbone to support various concurrent tasks: (i) real-time load balancing, (ii) optimal DER dispatch, and (iii) predictive assessment of stability margins. The generative capabilities of M3T FedFMs can further transform them from a purely prescriptive control engine into a strategic decision-support tool. For example, given current grid conditions, market forecasts, and asset statuses, the model can generate control policy operational blueprints containing multiple feasible dispatch strategies, each annotated with performance trade-offs such as cost efficiency, carbon footprint, and resilience under potential contingencies. These operational blueprints can be expressed in operator-friendly natural language, accompanied by data-driven rationale, enabling human-in-the-loop validation.

⋆ To enable the above applications, we next present a series of insights into the integration of M3T FedFMs within the smart grid ecosystem, with a particular focus on model training/fine-tuning and model utilization. Each insight is concluded with a set of targeted research questions, aiming to illuminate promising directions in this largely unexplored space.

*C. Federated Deployment Across Grid Hierarchies*

Modern smart grids are organized in hierarchical layers, ranging from low-level components such as substations and microgrids to higher-level entities like regional control centers and transmission system operators (see Fig. 2). These layers operate semi-independently, often managed by different organizations or vendors, and may differ in terms of their collected data modalities, and operational goals. Deploying M3T FedFMs within this setting requires a shift from conventional FL architectures considered in the M3T FedFM literature, which typically assume flat, direct node-to-server communication. In particular, smart grid deployments must support M3T FedFMs across structured hierarchies, where node model updates may be aggregated at intermediate levels (e.g., within microgrids or control regions) before contributing to higher-level global models. This hierarchical model training and aggregation raises several open questions, such as: (1) How should locally trained/fine-tuned M3T FMs be aggregated across tiers with varying data quality? (2) How should training/fine-tuning and aggregation timing be managed across layers?

*D. Personalization and Adaptability*

M3T FedFMs support model personalization by enabling each node, such as a substation or microgrid controller, to selectively fine-tune components of the model based on its own data distributions and operational priorities. For instance, a coastal utility may require models that are sensitive to wind- and salt-induced anomalies, whereas an urban operator may focus on congestion prediction and equipment aging under dense load conditions. Model personalization can be implemented through several mechanisms. Nodes may fine-tune task-specific heads or modality-specific encoders to better align with their localized inputs and objectives. Also, lightweight adaptation techniques, such as LoRA, prompt tuning, or adapter modules, enable resource-efficient personalization by updating only a small set of parameters while keeping the core model frozen. Some nodes may even maintain personalized branches within otherwise shared model components to allow partial divergence without disrupting global cohesion. Furthermore, the modularity of M3T FMs facilitates continual learning, allowing nodes to incrementally adapt to changes in consumption trends, seasonal dynamics, and infrastructure upgrades without retraining from scratch.

However, introducing model personalization in the M3T FedFM settings also poses new challenges. Model personalization can lead to model divergence, where overfitting to local data degrades global model performance. It also raises questions, such as: (1) How to balance global model generalization with local model relevance through efficient combination of the global and local models? (2) Which model personalization technique should be used given the downstream tasks of interest across the grid network hierarchy? Additionally, integrating continual learning with federated updates introduces temporal coordination complexities with research questions such as: When should a node update its model locally, and when should it re-synchronize with the global model?

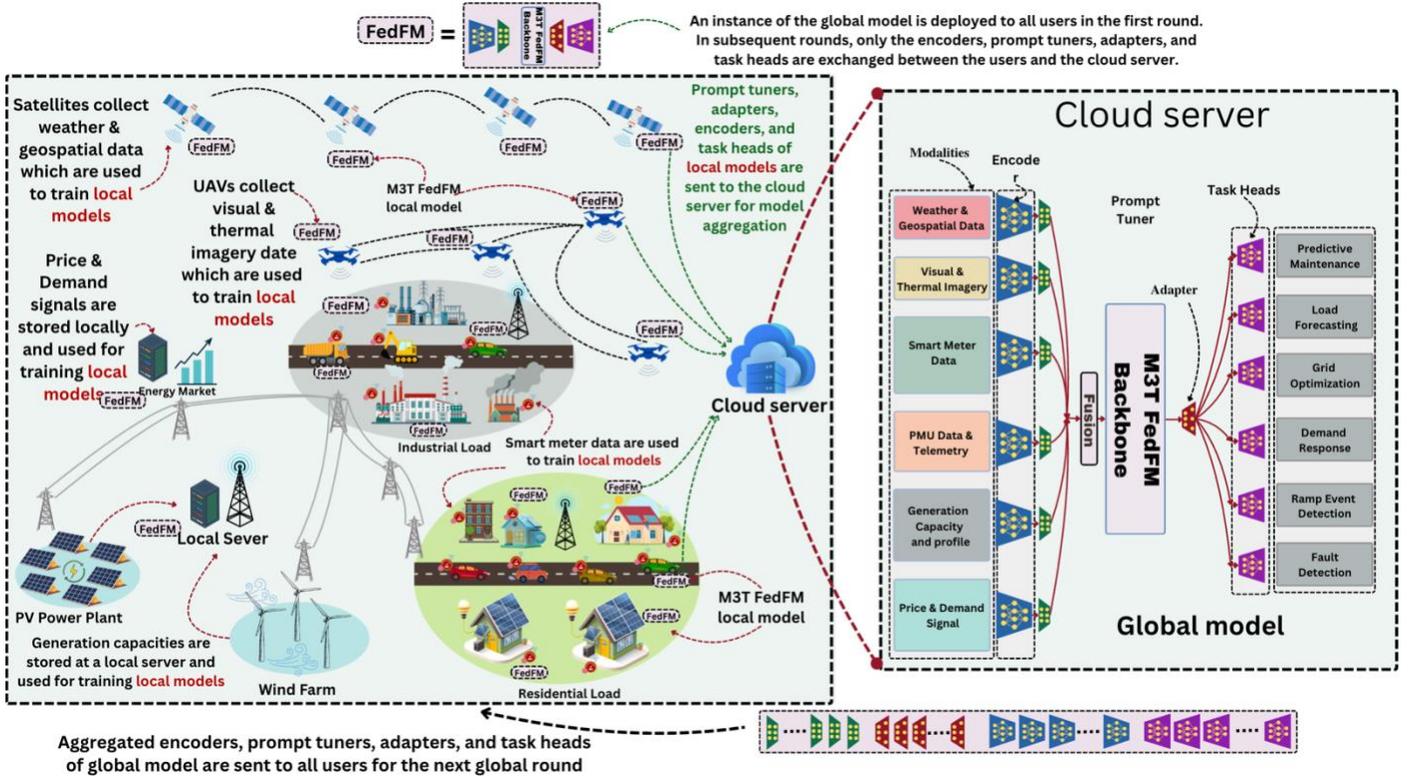

Fig. 2. Left: A smart grid system running M3T FedFM, where diverse users collect multi-modal data (e.g., weather, imagery, telemetry) to train/fine-tune local M3T FMs and engage in global aggregations with a server. Right: A cloud server hosting the global M3T FedFM with modality-specific encoders, prompt tuners, and adapters for fine-tuning. The model possesses task-specific heads for grid applications such as load forecasting and fault detection. In each round, depending on the fine-tuning technique used, only encoders, prompt tuners, adapters, and task heads are exchanged for aggregation and redistributed to all users.

*E. FedFMs Synergizing with Physics-based Tools*

We next shed light on a nuanced aspect of M3T FedFMs that intersects ML with the physical laws governing power system behavior. The use of M3T FedFMs in various applications for smart grids rests on an ambitious assumption: that these models can learn the underlying physical constraints of the grid, such as the nonlinear power flow equations derived from Kirchhoff's laws. Despite the theoretical power of neural networks as universal approximators, enforcing complex nonlinear constraints, especially equality constraints, of large-scale systems (involving thousands or tens of thousands of nodes) remains practically intractable using existing soft or hard constraint methods.

To address this, we envision a hybrid learning paradigm that tightly integrates M3T FedFMs with physics-based solvers. In this framework, the strength of the underlying M3T FMs trained at nodes is leveraged in capturing context and relationships to provide intelligent recommendations on what simulation scenarios to analyze and how to configure physical solvers (e.g., what demand or contingency scenarios to prioritize and how to configure physical solvers). The physics-based solvers (e.g., simulators, optimal power flow solvers) then solve the constrained nonlinear programming problems to deliver accurate operational insights and optimal decisions. This untapped symbiotic design respects the divide between what M3T FMs and M3T FedFMs can learn and what must be governed by the domain physics. This prompts two important research questions: (1) How can we design reliable and interpretable interfaces between generative M3T FMs (trained via M3T FedFMs) and physics-based solvers to ensure physically consistent decision-making? (2) Which classes of physical constraints are best embedded as inductive biases within M3T FedFMs, and which should be explicitly enforced by external solvers?

IV. SMART GRIDS FOR M3T FEDFMS: INFRASTRUCTURE CONSTRAINTS AND DESIGN CRITERIA

In this section, we examine the system-level requirements that influence the design, training, and deployment of M3T FedFMs in smart grid environments. Each of the following parts is concluded with a set of research questions, aiming to identify relevant research directions.

*A. Energy Burdens of M3T FedFM Training at Scale: Grid-Aware Scheduling, Load Coordination, and Beyond*

Data centers account for a significant share of global electricity usage, a footprint that is expected to grow sharply with the continued adoption of energy-intensive workloads such as training LLMs, M3T FMs, and more recently, M3T FedFMs. In particular, while M3T FedFMs may shift the model fine-tuning to distributed edge or regional nodes, the orchestration, aggregation, personalization, and model pre-training phases of M3T FedFMs still fundamentally depend on large-scale data center infrastructures. These operations require massive GPU/CPU clusters and large memory footprints, thereby intensifying the demand for both electricity and water, especially for cooling systems that prevent overheating and hardware failures in data centers. Crucially, the rising load from the continual training of these large ML models can introduce new operational risks to the grid: due to their higher workloads, data centers may become more sensitive to voltage disturbances and frequently disconnect from the grid during such events, triggering transitions to backup generators. These abrupt transitions can force excess electricity back into the grid, compelling operators to rapidly curtail power generation to prevent instability or cascading outages. Recent reports indicate that indeed such *near-miss* events have become more frequent as data center usage continues to grow. In this context, M3T FedFMs are not only consumers/providers of grid intelligence as detailed in the previous section, *they are also contributors to its load volatility and operational stress*.

To address this dual synergy, one promising approach lies in grid-aware scheduling and load coordination for M3T FedFMs, which aim to align the timing and location of compute operations with real-time grid conditions and long-term sustainability goals. For example, training M3T FedFMs during off-peak hours or periods of renewable energy surplus can reduce grid stress and operational costs. Furthermore, efficient workload orchestration across globally distributed data centers enables shifting compute tasks of M3T FedFMs to regions with favorable grid conditions, whether in terms of carbon intensity, electricity pricing, or infrastructure resilience. The design of such operations calls for addressing two untapped research questions: (1) How can we design grid scheduling and workload orchestration techniques that can optimize M3T FedFM training across distributed data centers based on real-time grid conditions, carbon intensity, and energy cost? (2) What role can M3T FedFMs themselves play in predicting, mitigating, or coordinating grid stress events triggered by their own training and aggregation workloads?

### B. Communication Constraints

The performance of smart grid applications is highly sensitive to the quality of the underlying communication infrastructure. Key performance indicators (KPIs), including throughput, latency, jitter, and packet loss, directly impact the reliability and effectiveness of critical grid functionalities such as demand response, fault detection, protection coordination, and DER dispatch. For instance, excessive delay or packet loss in control or protection loops can lead to instability or late actuation, posing serious risks to system safety and resilience. These challenges are particularly acute at the grid edge, where field devices such as smart meters, inverters, or sensors often rely on constrained communication technologies (e.g., Zigbee, LoRaWAN, LTE Cat-M), which inherently trade off between range, data rate, and power consumption. These challenges become even more pronounced with the deployment of M3T FedFMs over smart grids. These models require iterative, often asynchronous, communication between edge nodes and coordination servers for the exchange of model updates, gradient information, or control signals for aggregation and refinement. As a result, M3T FedFM deployment introduces a communication workload, potentially competing with mission-critical grid control and monitoring tasks that rely on the same resource-constrained communication infrastructure: for example, fine-tuning an M3T FedFM for DER management task while DERs simultaneously issue real-time actuation commands through the same wireless uplink channel that is used for transmitting the fine-tuned parameters.

At the same time, emerging communication technologies offer a promising solution for mitigating these resource contention and coordination challenges. The rollout of 5G and the anticipated capabilities of 6G networks provide native support for ultra-reliable low-latency communication (URLLC), massive machine-type communication (mMTC), and enhanced mobile broadband (eMBB), together enabling faster, more scalable, and quality-of-service (QoS)-flexible communication infrastructures. In parallel, the rise of Open Radio Access Network (O-RAN) architectures introduces programmable, open interfaces and ML-native modules directly within the RAN stack, which provides a unique opportunity to enforce dynamic prioritization and isolation between grid-critical control traffic (e.g., protection signals, DER commands) and M3T FedFM-centric flows (e.g., model parameter uploads and aggregation signals). This evolving landscape opens up several fundamental questions: (1) How can we design adaptive communication frameworks that coordinate M3T FedFM training traffic with real-time grid control traffic over shared wireless links, ensuring reliable model convergence for M3T FedFMs without disrupting the operations of safety-critical grid applications? (2) How can O-RAN's programmable interfaces and ML-native modules be leveraged to dynamically isolate, prioritize, or reshape M3T FedFM-related traffic flows?

### C. Governance and Regulatory Barriers

The deployment of M3T FedFMs over smart grids introduces not only technical challenges but also complex governance and regulatory questions. These models are inherently multi-party, trained across data silos owned by utilities, regional grid operators, equipment manufacturers, and possibly governmental or third-party entities. This distributed ownership raises various concerns: Who owns the resulting model? Who is responsible for updating, maintaining, and validating it? And who holds liability when the model's recommendations lead to suboptimal or unsafe grid behavior? Further, the decentralized execution of M3T FedFMs may raise two critical concerns: trustworthiness and interpretability. First, stakeholders may be reluctant to trust or adopt a model whose training pipeline spans multiple entities with unknown data quality, varying regulatory obligations, or conflicting incentives. Second, the inherent complexity of M3T FedFMs, especially when trained over multimodal, multi-task datasets, makes it difficult to understand or explain their internal reasoning processes. This poses a barrier for compliance with grid-sector regulations that increasingly require model explainability and human-in-the-loop oversight, especially in contexts such as fault diagnosis, protection coordination, or load shedding. These barriers give rise to two pressing research questions: (1) How can we design governance frameworks for M3T FedFMs that ensure transparent model collaboration and ownership across grid stakeholders, while embedding market-based incentives (e.g., auctions) to reward their contributions of data, compute, and model updates? (2) What techniques can enhance the trustworthiness and interpretability of M3T FedFMs in safety-critical smart grid contexts, while respecting the privacy of participating entities?

## V. Conclusion

In this paper, we introduced a bidirectional framework bridging smart grids and M3T FedFMs. We demonstrated how M3T FedFMs offer a unified solution to leverage heterogeneous, distributed data for diverse grid tasks, ranging from forecasting to control, while preserving the privacy of the collected data across the grid edge. Conversely, we examined how smart grid constraints across energy, communication, and governance domains impact the design and deployment of these models. By identifying the key capabilities of M3T FedFMs, the infrastructure bottlenecks facing their deployment, and various open research questions, this work lays a foundation for a highly promising yet largely unexplored research direction at the intersection of power systems and M3T FedFMs.

## VI. Acknowledgment

The authors would like to thank Payam Abdisarabshali for the graphical design of the figures.

## VII. For Further Reading

**Seyyedali Hosseinalipour** (alipour@buffalo.edu) is currently an Assistant Professor with the EE Department at the University at Buffalo–SUNY, USA.

**Shimiao Li** (shimiaol@buffalo.edu) is currently an Assistant Professor with the EE Department at the University at Buffalo–SUNY, USA.

**Adedoyin Inaolaji** (ainaolaj@buffalo.edu) is currently an Assistant Professor with the EE Department at the University at Buffalo–SUNY, USA.

**Filippo Malandra** (filippom@buffalo.edu) is currently an Assistant Professor with the EE Department at the University at Buffalo–SUNY, USA.

**Luis Herrera** (lcherrer@buffalo.edu) is currently an Associate Professor with the EE Department at the University at Buffalo–SUNY, USA.

**Nicholas Mastronarde** (nmastron@buffalo.edu) is currently a Professor with the EE Department at the University at Buffalo–SUNY, USA.